\documentclass[aps,prb,preprint,superscriptaddress]{revtex4-2}

\usepackage{graphicx}
\usepackage{amsmath}
\usepackage{braket}
\usepackage{booktabs}
\usepackage{longtable}
\usepackage{xcolor}
\usepackage{siunitx}
\usepackage[version=4]{mhchem}
\usepackage[english]{babel}

\newcommand{\SiN}{Si$_3$N$_4$}
\newcommand{\SiO}{SiO$_2$}

\newcommand{\neff}{n_{\mathrm{eff}}}
\newcommand{\ngrp}{n_{\mathrm{g}}}

\begin{document}

\title{Stitch-Free, Diamond-Scribed Silicon Nitride Photonic Integrated Circuits for the Visible Band}

\author{Kishor Kumar Mandal$^{\dagger}$}
\affiliation{Laboratory of Optics of Quantum Materials, Department of Physics,
Indian Institute of Technology Bombay (IITB), Mumbai - 400076, India}

\author{Lekshmi Eswaramoorthy$^{\dagger}$}
\affiliation{Laboratory of Optics of Quantum Materials, Department of Physics,
Indian Institute of Technology Bombay (IITB), Mumbai - 400076, India}
\affiliation{IITB-Monash Research Academy, Indian Institute of Technology Bombay,
Mumbai - 400076, India}

\author{Parul Sharma$^{\dagger}$}
\affiliation{Laboratory of Optics of Quantum Materials, Department of Physics,
Indian Institute of Technology Bombay (IITB), Mumbai - 400076, India}

\author{Anuj Kumar Singh}
\affiliation{Laboratory of Optics of Quantum Materials, Department of Physics,
Indian Institute of Technology Bombay (IITB), Mumbai - 400076, India}

\author{Brijesh Kumar}
\affiliation{Laboratory of Optics of Quantum Materials, Department of Physics, Indian Institute of Technology Bombay (IITB), Mumbai - 400076, India}

\author{Tanmay Gupta}
\affiliation{OptixLog (Coupler Inc.), San Francisco, CA-94107, USA}

\author{Venu Gopal Achanta$^{*}$}
\affiliation{Department of Condensed Matter Physics and Material Science,
Tata Institute of Fundamental Research, Homi Bhabha Road, Mumbai, 400005 India}

\author{Anshuman Kumar$^{*}$}
\affiliation{Laboratory of Optics of Quantum Materials, Department of Physics,
Indian Institute of Technology Bombay (IITB), Mumbai - 400076, India}
\affiliation{Centre of Excellence in Quantum Information, Computation, Science and Technology (QuICST),
Indian Institute of Technology Bombay, Mumbai - 400076, India}

\email[Authors to whom correspondence should be addressed: ]{achanta@tifr.res.in, anshuman.kumar@iitb.ac.in}
\thanks{$^{\dagger}$These authors contributed equally to this work.}

\begin{abstract}
Silicon nitride photonic integrated circuits for the visible band are
conventionally built with a buried oxide overcladding and singulated with
wafer-scale tooling, constraints that preclude evanescent access to the guided mode for externally integrated emitters. We report a PECVD \ce{Si3N4} platform designed around an air-clad waveguide whose evanescent field remains accessible along the full device length. Two process elements make this geometry practical at chip scale. Fixed-beam moving-stage electron-beam lithography writes {\SI{500}{\nano\meter}} single-mode waveguides as one continuous exposure across the \SI{5}{\milli\meter} chip, removing write-field stitching which, given the $\sigma^{2}/d^{4}$ scaling of sidewall scattering in this high-confinement geometry at \SI{635}{\nano\meter}, would otherwise dominate the loss budget. Chip singulation is performed by pen-type diamond scribing along lithographically patterned markers registered to in-plane direction, cleaving the Si(100) substrate to yield end-facets within \SI{2}{\degree} of normal at \SI{80}{\percent} yield. Structural characterization by scanning electron microscopy confirms stitch-free waveguide geometry and undamaged, near-vertical scribed facets; light is coupled end-fire into fabricated devices and guided to a microring with evanescent bus-to-ring coupling confirmed by scattering imaging, and a sidewall-roughness-dependent scattering-loss model indicates that loss remains low in the roughness regime consistent with the observed facet and sidewall quality. Building on the intrinsic emitter--resonator coupling demonstrated in \cite{mandal2024emission}, this platform extends monolithic \ce{Si3N4} photonics toward scalable visible-to-near-infrared quantum and classical circuits.
\end{abstract}

\maketitle

\vspace{4pt}
\noindent\rule{\linewidth}{0.4pt}
\vspace{8pt}

\section{Introduction}

\label{sec:intro}
Silicon nitride is transparent across the visible and near-infrared and is
compatible with standard CMOS processing, which has established it as a
principal platform for integrated photonics at wavelengths where silicon --
the workhorse material of mature, telecom-band silicon photonics -- is
opaque~\cite{Moss2013_SiN_PICs,Buzaverov2024_SiN_review,Munoz2017}. Its function in quantum photonic architectures is no longer restricted to passive routing: solid-state quantum light sources built around atomic-scale defects and point emitters are increasingly regarded as a route toward scalable, chip-integrated single- and entangled-photon technologies~\cite{singh2026perspective, zelaya2025chip}. \ce{Si3N4} has been shown to host optically active intrinsic defects of this kind: in recent work we coupled such emitters to whispering-gallery modes of monolithically integrated \ce{Si3N4} microrings, employing a subwavelength notch in the ring rim to enhance the collection of cavity-coupled photoluminescence at room temperature~\cite{mandal2024emission}. A single nitride layer can thus serve simultaneously as the photonic and the emitter host.

The extension of this result from an isolated resonator to a circuit requires optical access to the guided mode along the length of the device. Intrinsic \ce{Si3N4} emitters are distributed throughout the core, but the complementary route to visible-band on-chip emission - evanescent coupling to a two-dimensional material such as hexagonal boron nitride transferred onto the waveguide surface~\cite{singh2025plasmonic} - requires that the mode remain accessible from above. Photonic-chip architectures coupling guided light to two-dimensional semiconductor excitons have already been demonstrated for on-chip valley and exciton routing~\cite{kumar2023photonic, kumar2023universal}, and the resulting dark- and bright-exciton emission has been shown to depend sensitively on the local photonic environment through Purcell factor engineering~\cite{eswaramoorthy2022purcell} -- underscoring the need for evanescent, rather than buried, access to the mode. A buried oxide overcladding, employed in most low-loss \ce{Si3N4} platforms, precludes such access. The air-clad geometry is therefore adopted here as a design constraint, and the remainder of the process flow is developed subject to it.

The air-clad geometry imposes correspondingly stringent requirements on
sidewall quality. At $\lambda \sim \SI{635}{\nano\meter}$ with a \SI{500}{\nano\meter}~$\times$~\SI{300}{\nano\meter} core, sidewall scattering
loss scales as $\sigma^{2}/d^{4}$, where $\sigma$ is the RMS roughness and $d$ the waveguide half-width~\cite{payne1994,roberts2022}. This geometry is
therefore two to three orders of magnitude more sensitive to roughness than the telecom-band, micron-scale waveguides -- including mature silicon photonics platforms operating near \SI{1550}{\nano\meter} -- in which sub-\si{\decibel\per\meter} propagation losses have been demonstrated. Scattering from fabrication-induced imperfection, rather than material absorption, is consequently expected to dominate the loss budget. The exposed upper surface is not the principal contributor: for a comparable high-confinement cross-section, as-deposited surface roughness has been shown to contribute approximately two orders of magnitude less loss than plasma-etched sidewall roughness~\cite{ljubotina2024}. Two other extrinsic mechanisms are more consequential. The first is write-field stitching in electron-beam lithography (EBL), which introduces lateral displacements of \SIrange{10}{50}{\nano\meter} at each field boundary; over a centimeter-scale waveguide traversing multiple fields, the accumulated contribution may exceed the intrinsic propagation loss. The second is the quality of the chip end-facet.

Efficient fiber-to-chip edge coupling requires end-facets that are smooth,
vertical, and perpendicular to the propagation axis. Surface roughness
introduces scattering and reflection, while angular deviation of the facet
plane displaces the output beam from the fiber acceptance
cone~\cite{HighEffEdgeCoupling2025,Gow2024_MechanicalDicing}. The established route to optical-grade facets in \ce{Si3N4} is automated blade dicing followed by chemical-mechanical polishing (which requires a relatively complex, costly, and time-intensive fabrication process), since dicing alone produces a subsurface damage zone extending several micrometers beneath the cut edge and facet roughness of tens to hundreds of nanometers RMS~\cite{HighEffEdgeCoupling2025, Gow2024_MechanicalDicing}. Stealth dicing and
etch-defined chip edges offer alternative routes, but require a dedicated laser tool or additional lithography and etch steps at the chip perimeter,
respectively. Focused-ion-beam polishing has separately been used to recover high-\textit{Q} surface finish on resonator facets~\cite{eswaramoorthy2026fib}, but as a serial, low-throughput technique it is likewise poorly suited to wafer- or chip-scale singulation. In the present work, blade dicing of the \ce{Si}/\ce{SiO2}/\ce{Si3N4} stack yielded wavy, chipped end-faces unsuitable
for reproducible coupling, which we attribute to insufficient optimization of
the dicing parameters for this film stack rather than to a limitation of the
technique itself~\cite{Gow2024_MechanicalDicing}; the polishing step required to recover such facets was not available in the present cleanroom environment.

We report here a \ce{Si3N4} platform addressing both extrinsic mechanisms
without recourse to additional capital equipment. Fixed-beam moving-stage (FBMS)
electron-beam lithography is used to expose width $\sim 500$ nm single-mode waveguides as a single continuous scan across the full \SI{5}{\milli\meter} chip length, eliminating write-field boundaries. A layout rule confines all FBMS-to-write-field transitions to the inverse-taper region, such that residual registration offsets coincide with an
already-varying local mode profile rather than with an abrupt discontinuity.
Chips are singulated by pen-type diamond scribing along lithographically
patterned markers and cleaving of the Si(100) substrate along the scratched line to yield end-facets
within \SI{2}{\degree} of the substrate normal at a yield of
\SI{80}{\percent}. Supporting process elements are described in detail
sufficient for transfer to other facilities: a titanium hard mask deposited by electron-beam evaporation, which avoids the metal crowning associated with
sputter deposition at sub-micron feature edges, and a pulsed sonication lift-off protocol that removes the residues responsible for point scattering along
extended waveguides. In combination with the air cladding, these elements
provide a continuous interaction length that is optically accessible along its full extent and edge-coupled at both facets.

\section{Methods}
\label{sec:Methods}
 
\subsection{SiN Photonic Chip Modelling}
\label{ssec:modal}
\begin{figure}
    \centering
    \includegraphics[width=0.95\columnwidth]{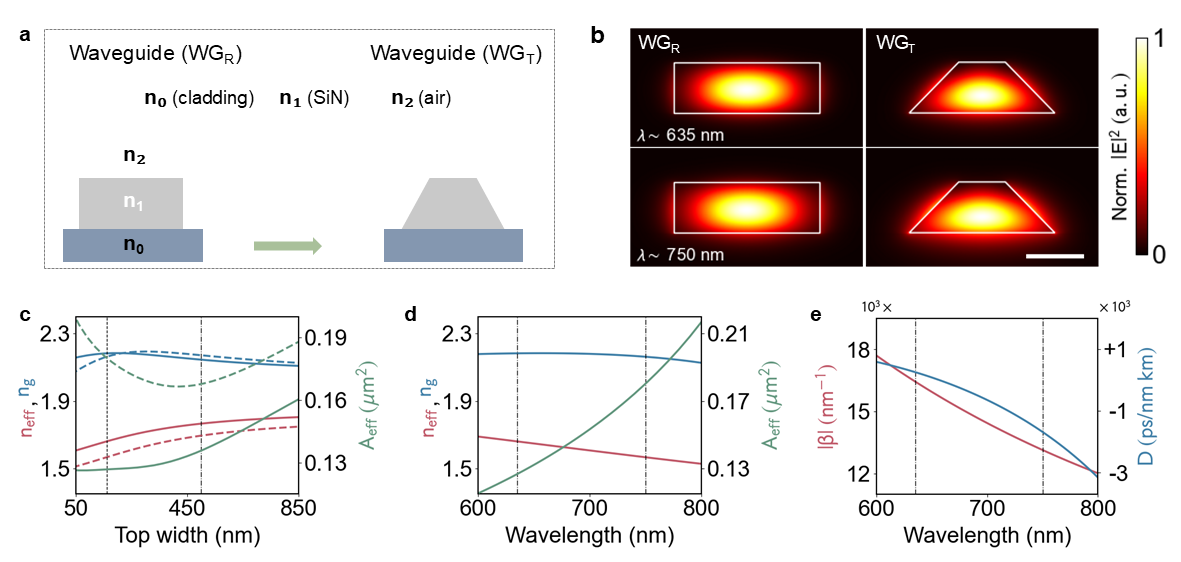}
    \caption{(a) Schematic of the waveguide cross-section (rectangular and trapezoidal); (b) FDE-computed TE$_{00}$ mode profiles at 635 and 750\,nm  wavelength for both the cross-sections; (c) $\neff$, $\ngrp$ and $A_{eff}$ vs.\ waveguide width at  635 and 750\,nm; (d) $\neff$, $\ngrp$ and $A_{eff}$ vs.\ wavelength; (e) Propagation constant and dispersion vs wavelength.}
    \label{Fig:Fig1}
\end{figure}

 \textcolor{black}{The waveguide geometry is designed to efficiently guide the fundamental transverse electric mode (TE$_{00}$) at the target wavelengths of 635 (for excitation) and 750~nm (for hybrid integration of emitter to interact with cavity), corresponding to the spectral wavelengths WSe$_2$. We consider a silicon nitride \SiN\ core deposited a-top the SiO$_2$ bottom cladding layer, with a core thickness of 300~nm deposited by plasma-enhanced chemical vapor deposition (PECVD). Both rectangular (isolated core mode type) and fabrication-relevant trapezoidal cross-sections (provides strong modal overlap at the coupling junction)  are considered to evaluate the influence of sidewall geometry on the guided mode, as shown in the Fig.~\ref{Fig:Fig1}(a). Fig.~\ref{Fig:Fig1}(b) compares the simulated modal profiles for the two cross-sections at 635 and 750~nm wavelength. The normalized electric-field intensity, $|E|^2$, shows that the fundamental TE-like mode remains well confined within the \SiN\ core for both wavelengths.}
 
 \textcolor{black}{To find suitable lateral waveguide dimensions, the effective index ($n_{\mathrm{eff}}$), group index ($n_{\mathrm{g}}$), and effective mode area ($A_{\mathrm{eff}}$) are evaluated as functions of waveguide top width (of trapezoidal cross-section) and wavelength using a finite-difference eigenmode (FDE) solver (Ansys/Lumerical MODE solver), as shown in Fig.~\ref{Fig:Fig1}(c,d). These parameters are evaluated over the experimentally accessible width of approximately 500~nm (at bottom), corresponding to an effective width of 330~nm (achieved after etching) at half the core-layer thickness, to provide strong evanescent coupling. The wavelength-dependent propagation characteristics are further evaluated through the propagation constant, $\beta$, and the dispersion parameter, $D$, as shown in Fig.~\ref{Fig:Fig1}(e). The calculated $\beta(\lambda)$ describes the phase evolution of the guided mode, whereas $D$ quantifies the wavelength dependence of the group velocity and therefore the chromatic dispersion of the waveguide.}

\begin{figure}
    \centering
    \includegraphics[width=0.9\columnwidth]{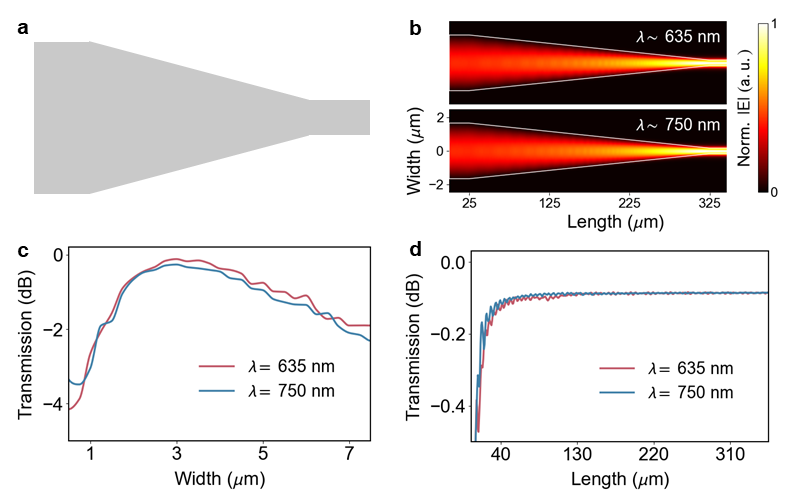}
    \caption{(a) Schematic of the taper geometry; (b) electric-field evolution along the optimized adiabatic taper length; (c) FDTD-computed coupling efficiency versus tip width for a lensed fiber with an MFD of $2~\mu\mathrm{m}$; (d) taper-length optimization for a fixed taper mouth width, simulated using the EME solver.}

    \label{fig:Fig2}
\end{figure}

\label{ssec:taper}
\textcolor{black}{The TE$_{00}$ mode of trapezoidal cross-section \SiN\ waveguide is substantially smaller than the mode field of a lensed optical fiber, resulting in significant mode mismatch at the chip facet. To reduce the resulting insertion loss, an inverse taper is incorporated at each chip edge, where the waveguide width is gradually reduced toward the facet, allowing the guided mode to expand and improve its overlap with the fiber mode. Lensed fibers with an MFD of approximately $2~\mu$m have been demonstrated for \SiN\ waveguides in the visible-to-near-infrared regime \cite{Sanna2024}. As shown in Fig.~\ref{fig:Fig2}(a,b), the inverse taper provides gradual mode expansion for both $\lambda=635$ and $750$~nm. For adiabatic mode evolution, the local taper angle must remain sufficiently small compared with the modal separation between the fundamental and the nearest higher-order mode \cite{Liang2022}. This condition can be expressed as
\begin{equation}
\tan\theta \ll
\frac{n_{\mathrm{eff}}^{(0)}-n_{\mathrm{eff}}^{(1)}}{\bar{n}},
\label{eq:adiabatic}
\end{equation}
where $n_{\mathrm{eff}}^{(0)}$ and $n_{\mathrm{eff}}^{(1)}$ are the effective indices of the fundamental and first-order modes, respectively, and $\bar n$ is their mean effective index. The taper width and length are optimized using finite-difference time-domain (FDTD) and eigenmode expansion (EME) simulations. Fig.~\ref{fig:Fig2}(c) shows the transmission as a function of taper width, exhibiting a broad maximum $\sim 3~\mu\mathrm{m}$ for both wavelengths, while Fig.~\ref{fig:Fig2}(d) shows that the transmission approaches saturation for taper lengths above approximately $100~\mu$m. The selected taper geometry therefore provides efficient and broadband mode expansion at both target wavelengths. The EBL resolution and the aspect ratio sustainable during reactive ion etching (RIE) without pattern collapse constrain the minimum achievable tip width. Critically, the layout requires that the FBMS-to-write-field boundary (refer to fabrication section) always falls within the taper region, so that any residual positional offset at this boundary is distributed over the gradually evolving local mode profile rather than constituting an abrupt
scattering discontinuity.}

\begin{figure}
    \centering
    \includegraphics[width=\columnwidth]{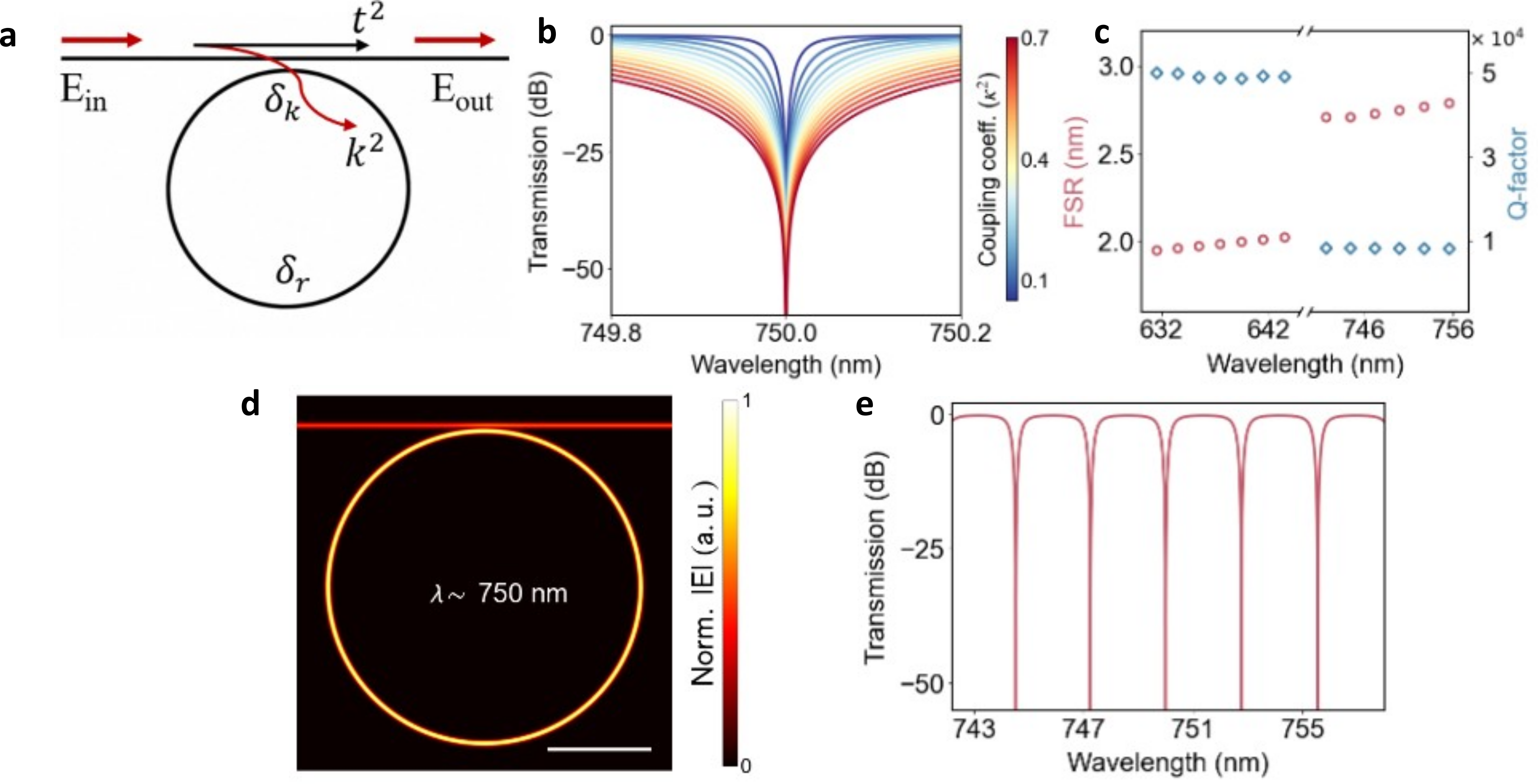}
    \caption{Spectral characteristics of the ring resonator system (a) Schematic illustration of a waveguide coupled to a ring resonator. (b) Calculated transmission spectra for different coupling coefficients, showing the evolution of the resonance lineshape. (c) Wavelength dependence of the free spectral range (FSR) and \(Q\)-factor. (d) Normalized electric-field intensity distribution of the resonant mode at \(\lambda\sim750\) nm. (e) Transmission spectrum of the microring resonator around the emitter emission wavelength of $\sim 750$~nm. 
    }
    \label{fig:ring_theory}
\end{figure}

\textcolor{black}{From coupled-mode theory (CMT), the co-directional coupling between the bus waveguide and the microring curved waveguide  can be analyzed by solving the forward-propagating modes (having propagation constant of modes $\beta_{i,j}=2\pi n_{\mathrm{eff},i,j}/\lambda$) within the two-waveguide at the coupling regime. The interaction results from the perturbation and spatial overlapping of the uncoupled waveguide modes, that generate the characteristic coupling coefficient between modes \(i\) and \(j\) \cite{van2016optical, little1997microring, yuan2023mechanisms},
\begin{equation}
K_{i,j}
=
\frac{\omega}{4}
\int
\left(\varepsilon-\varepsilon_j\right)
\mathbf{e}^{(i)}
\cdot
\mathbf{e}^{(j)}
\,dx\,dy
\end{equation}
\\
\textcolor{black}{Where, $\omega$ is the angular frequency of resonating modes, \(\varepsilon-\varepsilon_i\) represents the dielectric perturbation function, and $\mathbf{e}^{(i)}(x,y)$ and $\mathbf{e}^{(j)}(x,y)$ are the modal cross-sectional electric-field distributions. The corresponding input and output field amplitudes are related through the scattering matrix $\mathbf{S}$ (expressed as $\mathbf{E}_{\mathrm{out}}=\mathbf{S} \mathbf{E}_{\mathrm{in}}$}), 
\begin{equation}
\mathbf{S}=
\begin{bmatrix}
\tau & -j\kappa \\
-j\kappa & \tau
\end{bmatrix}
\end{equation}
where $\tau$ is the self-coupling field coefficient and $\kappa$ is the cross-coupling field coefficient of the coupler and for a lossless  coupler system, it satisfies the condition $|\tau|^2+|\kappa|^2=1$. This coupled-waveguide geometry -- a straight bus addressing a ring resonator
through a single coupling junction of coefficient $\kappa$ -- is shown
schematically in Fig.~\ref{fig:ring_theory}(a). The microring optical path ($n_{\mathrm{eff}}L$) provides continuous feedback within the microcavity by recirculating the selectively coupled resonanting modes, resulting in a periodic resonance spectrum with a free spectral range (FSR), $\Delta\lambda$, governed by the resonance condition $m\lambda=n_{\mathrm{eff}}L$; where, \(m\) is the integer resonance order. The transmission spectrum of the bus-coupled-microring resonator is expressed as}

\begin{equation}
T=\left|\frac{E_{\mathrm{out}}}{E_{\mathrm{in}}}\right|^2=\left|\frac{\tau-ae^{-j\phi}}{1-\tau ae^{-j\phi}}\right|^2
\label{eq:transmission}
\end{equation}

Fig.~\ref{fig:ring_theory}(b) shows the transmission spectra obtained from
Eq.~\eqref{eq:transmission} as $\kappa^{2}$ is swept from the under-coupled
toward the over-coupled regime, illustrating how the resonance depth and
linewidth evolve with coupling strength for a representative resonance near
\SI{750}{\nano\meter}. \textcolor{black}{Here, $a$ is the microcavity round-trip field transmission coefficient and is defined in terms of the propagation loss coefficient of the feedback loop as $a^2=e^{-\delta_r}$; $\phi$ is the corresponding round-trip phase; and $\tau$ is related to the coupling loss coefficient of the microring at the coupling junction by $\tau^2=e^{-\delta_{\kappa}}$. The key figures of merit of the resonance spectrum of an optical cavity are the loaded quality factor, $Q=\lambda/\delta\lambda$ and, in the time domain, in terms of the photon lifetime as $Q=\omega\tau_{\mathrm{ph}}$
; the resonance linewidth, defined by the full width at half maximum (FWHM), $\delta\lambda$; the finesse, $F$, which represents the resonance linewidth relative to the FSR; and the extinction ratio (ER), which characterizes the depth of the resonance dip,}

\begin{equation}
F=\frac{\Delta\lambda}{\delta\lambda}\approx\frac{2\pi}{\delta_{\kappa}+\delta_r}, \qquad
ER\approx\left|\frac{\delta_{\kappa}+\delta_r}{\delta_{\kappa}-\delta_r}\right|^2
\end{equation}

Over a broad wavelength range, the FSR is more accurately represented by accounting for waveguide dispersion through the group index,

\begin{equation}
n_g=n_{\mathrm{eff}}-\lambda\frac{dn_{\mathrm{eff}}}{d\lambda}
\end{equation}

and the corresponding FSR is given by

\begin{equation}
FSR=\frac{\lambda^2}{n_gL}
\label{eq:fsr_dispersion}
\end{equation}

The resulting wavelength dependence of the FSR and loaded $Q$-factor,
evaluated separately in the vicinity of the \SI{635}{\nano\meter} and
\SI{750}{\nano\meter} design wavelengths, is summarized in
Fig.~\ref{fig:ring_theory}(c). The larger FSR and lower $Q$ obtained near
\SI{635}{\nano\meter} relative to \SI{750}{\nano\meter} follow directly from
the stronger sidewall-scattering sensitivity at shorter wavelength discussed
in Section~\ref{sec:intro} ($\sigma^{2}/d^{4}$ scaling). The significant reduction in the $Q$-factor caused by waveguide sidewall roughness can be correlated with the corresponding increase in propagation loss through the following relationship

\begin{equation}
Q=\frac{2\pi n_g}{\alpha\lambda}
=\frac{\lambda}{\alpha \cdot R \cdot FSR}
\label{eq:Q_factor}
\end{equation}

Applying this formalism to the ring geometry used in the fabricated devices,
  Fig.~\ref{fig:ring_theory}(d) shows the FDTD-simulated $\mathrm{TE}_{00}$ mode profile of the resonant ring waveguide at $\lambda \sim \SI{750}{\nano\meter}$; the field remains tightly confined to the ring core with negligible amplitude at the outer rim, consistent with the low-radiation-loss regime assumed above. Fig.~\ref{fig:ring_theory}(e) shows the corresponding simulated transmission spectrum of the bus-coupled microring over a wavelength window centered on this emitter emission wavelength, exhibiting a periodic series of resonance dips whose spacing matches the FSR obtained from Fig.~\ref{fig:ring_theory}(c).

Fig.~\ref{fig:ring_theory}(d) shows the whispering-gallery mode (WGM) of light coupled into the microring cavity at a wavelength around 750~nm, while Fig.~\ref{fig:ring_theory}(e) shows the transmission spectrum of the resonant mode with the corresponding FSR.

\subsection{Fabrication and chip singulation}
\label{ssec:fabrication}
The devices are fabricated on commercially available Si wafers with $\langle100\rangle$ surface orientation and a nominal thickness of $\sim500~\mu$m. Following RCA cleaning, a $2~\mu$m-thick \SiO\ layer is thermally grown to provide optical isolation from the Si substrate and a high-quality interface for the subsequent \SiN\ layer. A $300$-nm-thick \SiN\ film is then deposited by PECVD using NH$_3$ and SiH$_4$ precursors~\cite{Gu:14, 10.1063/5.0057881} under the process conditions established in our previous work~\cite{mandal2024emission}. The deposition is performed at an RF frequency of $13.56$~MHz, RF power of $400$~W, chamber pressure of $5$~Pa, and substrate temperature of $200^\circ$C~\cite{mandal2024emission}.

The photonic circuits are patterned by EBL because the waveguide widths of $\sim 500$~nm and bus-to-ring coupling gaps of $100$-$300$~nm require dimensional control beyond that readily provided by the available optical lithography infrastructure. The Si/SiO$_2$/SiN substrates are cleaned using acetone, IPA, and DI water, followed by UV-ozone treatment and dehydration baking. A multilayer PMMA resist stack is spin-coated at $4000$~r.p.m. and prebaked at $180^\circ$C for $2$~min~\cite{singh2024interplayplasmonicsstrainhexagonal}. EBL is performed using a Raith~150TWO system in FBMS-writing-mode, which enables continuous writing of the long waveguide structures, including circuits with lengths up to $\sim$~cm scale~\cite{10.1116/6.0002187}. The principal exposure parameters are an accelerating voltage of $30$~kV, $10~\mu$m aperture, beam current of $0.044$~nA, area dose of $500~\mu$C/cm$^{-2}$, working distance of $8$~mm. Proximity-effect correction is applied to compensate for electron-scattering-induced dose broadening in closely spaced and sub-micron features.

A particular fabrication consideration arises from the use of FBMS for millimeter-scale photonic circuits. The write-field boundaries and their positions are defined in the EBL job file prior to exposure, with alignment markers incorporated at the field boundaries to ensure accurate registration between FBMS and write-field pattern layers. A chip length of $7$~mm is selected to remain within the stable stage-travel and velocity range of the EBL system for the beam conditions employed. Because individual chips are substantially smaller than the standard wafer substrates for which EBL stages are designed, a dedicated chip holder was developed at the LOQM laboratory. The aluminium holder incorporates a recessed pocket for $5$-$10$~mm $\times$ $3$-$5$~mm chips, with the pocket depth matched to the substrate thickness so that the chip surface remains flush with the holder. The chip is secured using a thin, low-outgassing adhesive compatible with the EBL vacuum environment. This arrangement suppresses chip tilt and vibration during long FBMS scans and provides a mechanically stable platform for reproducible lithography.

Following EBL and development, the pattern is transferred into the \SiN\ using a $\sim90$-nm-thick Ti hard mask. Ti is deposited by electron-beam evaporation because its directional flux is favorable for lift-off of the sub-micron features and minimizes unwanted sidewall deposition. The Ti mask also provides substantially higher etch resistance than the organic PMMA resist during fluorine-based \SiN\ etching. After deposition, the unwanted Ti and underlying PMMA are removed by acetone lift-off, assisted by short ultrasonic pulses in fresh solvent and intermittent optical inspection. This step is particularly important for the long waveguides used here, where residual Ti or resist can introduce additional scattering and propagation loss. The resulting Ti patterns are inspected by optical microscopy and SEM before pattern transfer.

Pattern transfer from the Ti hard mask into the \SiN\ is performed using RIE with a fluorine-based plasma. The etch conditions are optimized to provide sufficient selectivity to the Ti mask while maintaining a predominantly vertical profile, which is important for accurately defining the waveguide widths and coupling gaps and for minimizing excess scattering and polarization-dependent loss. The process parameters are summarized in Table~\ref{tab:etch_params}.

\begin{table}[h!]
\centering
\caption{RIE process parameters for \SiN\ pattern transfer.}
\label{tab:etch_params}
\small
\begin{tabular}{lc}

    \textbf{Parameter} & \textbf{Value} \\
    
    Etch chemistry              & CHF$_3$/O$_2$ (98/2 sccm) \\
    Chamber pressure (Pa)    & 1  \\
    RF substrate bias (W)       & 150 \\
    \SiN\ etch rate (nm\,min$^{-1}$) & $\sim 28$ \\
    
\end{tabular}

\end{table}

Following RIE, the residual Ti hard mask is selectively removed using the wet-chemical procedure described in Ref.~\cite{mandal2024emission}. The etch duration is controlled to achieve complete Ti removal while preserving the underlying \SiN/\SiO\ stack. The completed photonic circuits are then singulated into individual chips for optical characterization.

\subsection{Chip layout and facet formation}

Reliable edge coupling is particularly important for the present devices because the optical signal is collected directly from the chip end-facet. The $5$-mm-long waveguides used in this work produce output powers of only $\sim30$~nW for an input power of $1$~mW at $635$~nm laser pumping. Consequently, losses introduced by facet roughness or angular misalignment can become comparable to the measurable transmitted signal. The singulation process must therefore produce facets that are sufficiently smooth and close to perpendicular to the waveguide propagation direction.

Two approaches were considered for chip singulation. Conventional diamond-blade dicing provides high throughput and dimensional accuracy but introduces mechanical damage, including chipping, microcracking, and subsurface damage near the cut edge. For the Si/\SiO/\SiN\ stack used here, initial dicing trials produced visibly rough and wavy facets with chipping near the waveguide layer. These facets were unsuitable for reproducible visible-wavelength fiber coupling without an additional polishing step. Although optical-grade facets can be obtained using a dicing chemical mechanical polishing (CMP) sequence, this approach requires specialized polishing equipment and consumables that were not readily accessible within the available fabrication infrastructure~\cite{Gow2024_MechanicalDicing, HighEffEdgeCoupling2025}, also, it can leave unwanted chemical residues and contamination particles on the top surface.

 \begin{figure}
    \centering
    \includegraphics[width= 0.7\columnwidth]{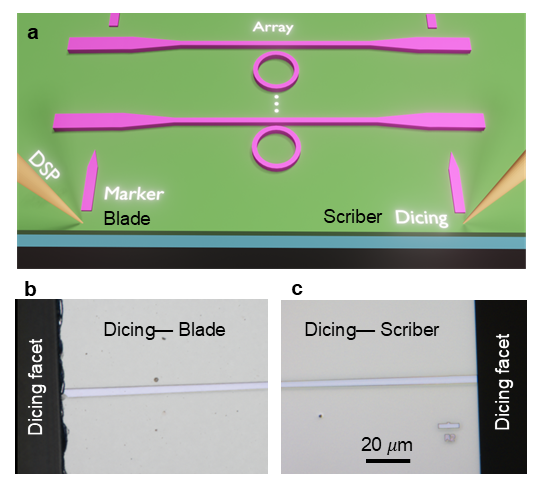}
    \caption{Comparison of wafer dicing methods (a) Schematic of the ring-waveguide array illustrating the positioning of the dicing regions and alignment marker. (b,c) Optical images of the resulting dicing facets obtained using b) direct blade dicing and c) scribing followed by dicing.}
    \label{fig:Fig6}
\end{figure} 

Other alternatives, including laser-based stealth dicing and lithographically defined etched edges, can provide improved facet quality but require either dedicated laser-dicing equipment or additional lithography and etching steps at the chip perimeter. These limitations motivated the development of a simple crystallographic singulation method based on diamond scribing and mechanical cleaving of the Si(100) substrate.

For a Si(100) substrate, the crystallographically preferred direction is used for cleaving the die sample. The EBL layout therefore incorporates dedicated scribing markers at the intended chip boundaries, allowing the diamond scriber to be aligned to the crystallographic direction under an optical microscope. The marker placement is particularly important because the chip dimensions are only a few millimeters, while the facet orientation must remain compatible with the $5$-mm waveguide axis. In the present process, the scriber can be aligned to the intended crystallographic direction with an accuracy of approximately $10~\mu$m. This marker-assisted procedure represents an additional fabrication step introduced in the present work and was not part of the singulation procedure reported previously~\cite{mandal2024emission}.

For singulation, the chip is positioned under an optical microscope and the diamond scriber is aligned with the designated scribing markers. A short $\sim1$-$2$~mm scribe is introduced with controlled pressure to initiate the fracture. The chip is subsequently placed on a clean, rigid surface and a symmetric bending force is applied across the scribe line, promoting crack propagation along the crystallographically preferred direction. The resulting facet is then subjected to brief lapping to remove residual surface debris and microscopic steps. SEM inspection is used to assess the facet orientation and surface morphology before optical coupling measurements.

The resulting cleaved facets exhibit a substantially cleaner and more uniform appearance than those obtained by the available dicing process. SEM measurements show near-vertical facets with an angular deviation of less than $2^\circ$ from the substrate normal, while optical inspection reveals the characteristic mirror-like appearance of the cleaved surface. Thus, the crystallographic cleaving approach provides a practical route to obtaining coupling-compatible facets without requiring dedicated dicing or CMP infrastructure.

The principal advantage of the developed method is that it combines low equipment requirements with rapid chip-level processing. A standard optical microscope and pen-type diamond scriber are sufficient, making the approach well suited to iterative PIC fabrication. The principal limitation is the dependence on scribing force, crystallographic alignment, substrate thickness, and local stress in the \SiN/\SiO/Si stack. In particular, the $\sim500~\mu$m substrate used here requires greater scribing force than thinner $\sim275~\mu$m substrates, although the thicker wafer provides improved mechanical robustness during the preceding wet-processing and EBL steps.

The procedure was evaluated over $N=20$ chips fabricated across multiple runs. Sixteen chips produced optically usable end-facets, corresponding to an overall singulation yield of 80$\%$.
The unsuccessful cleaves primarily resulted from deviation of the fracture trajectory from the intended crystallographic direction or uncontrolled fracture initiation near patterned regions. These results demonstrate that marker-assisted diamond scribing provides a practical and reproducible alternative to conventional mechanical dicing for the present long-waveguide \SiN\ PICs, while avoiding the additional CMP step otherwise required to obtain coupling-compatible facets.

\section{Results}
\label{sec:results}

\subsection{Facet and Taper Morphology}
\label{sec:results-sem}
\begin{figure}[htbp]
\centering
\includegraphics[width=0.95\linewidth]{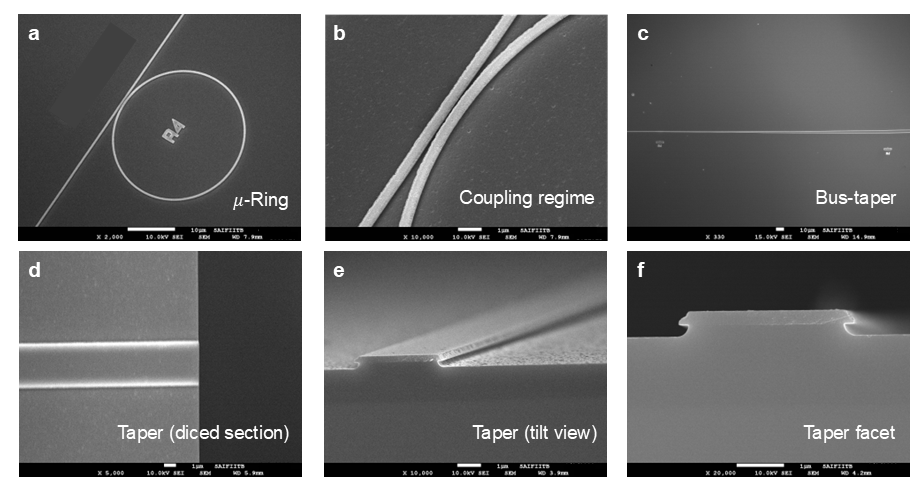}
\caption{Scanning electron micrographs of fabricated structures.
(a)~Microring resonator coupled to its bus waveguide. (b)~Bus-to-ring coupling
region. (c)~Inverse taper adjoining the single-mode bus waveguide.
(d)~Taper terminating at the singulated chip edge (plan view).
(e)~Tilted view of the taper sidewall profile. (f)~Taper end-facet after
diamond-scribe singulation and polishing.}
\label{fig:sem_facets}
\end{figure}

Structural verification of the FBMS-written waveguides, inverse tapers, and
scribed chip facets is obtained by scanning electron microscopy.
Figure~\ref{fig:sem_facets}(a) shows a representative microring coupled to its
bus waveguide, and Fig.~\ref{fig:sem_facets}(b) the bus-to-ring coupling region
at higher magnification; both waveguides present continuous, unbroken edges
over the imaged field, with no visible stitching artifact at the resolution of
the micrograph. Figure~\ref{fig:sem_facets}(c) shows the inverse taper adjoining
the single-mode bus section, and Figs.~\ref{fig:sem_facets}(d)--(f) show the
taper region after chip singulation: a plan view of the taper terminating at
the diced/scribed edge~\ref{fig:sem_facets}(d), a tilted view resolving the
taper sidewall profile~\ref{fig:sem_facets}(e), and the taper end-facet
itself~\ref{fig:sem_facets}(f). The facet in Fig.~\ref{fig:sem_facets}(f)
presents as flat and free of the chipping or waviness characteristic of
unoptimized blade dicing, consistent with
cleavage along a $\{100\}$ plane. These images provide structural, rather than
quantitative, evidence of process fidelity; they are considered together with
the roughness-dependent loss model of Section~\ref{sec:results-roughness} in
the Discussion.

\subsection{Optical Coupling and Transmission Measurements}
\label{sec:results-optical}

\begin{figure}[htbp]
\centering
\includegraphics[width=0.95\linewidth]{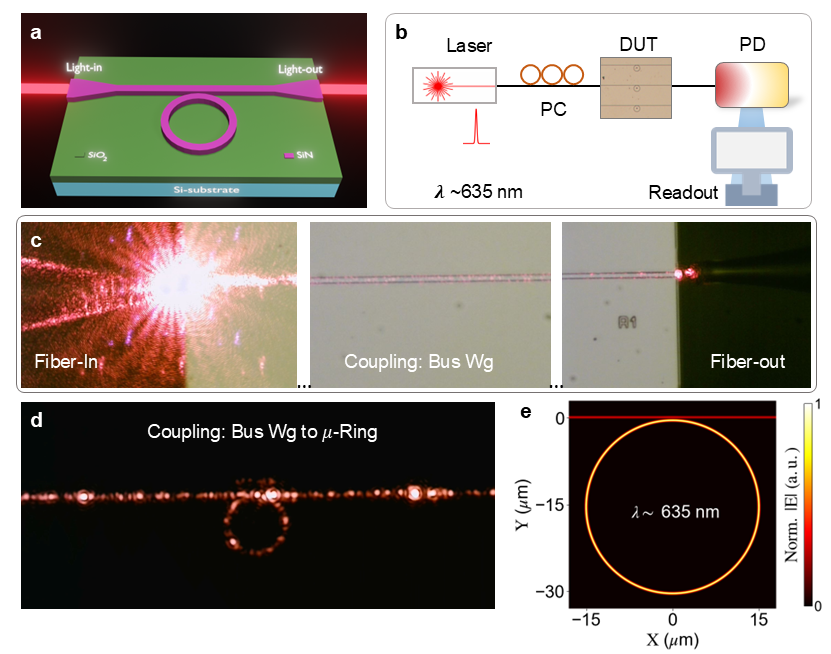}
\caption{Optical characterization setup and coupling verification.
(a)~Schematic of the device under test: a \ce{Si3N4} bus waveguide coupled to
a microring on a \ce{SiO2}/Si substrate, addressed by lensed fibers at each
chip facet. (b)~Block diagram of the home-built end-fire coupling setup: a
\SI{635}{\nano\meter} source, fiber polarization controller (PC), device under
test (DUT), photodetector (PD), and oscilloscope readout. (c)~Camera images
documenting the alignment sequence: light launched at the input facet
(Fiber-In), scattered light along the bus waveguide confirming guiding
(Coupling: Bus Wg), and light recollected at the output facet (Fiber-out).
(d)~Top-view scattering image showing evanescent coupling from the bus
waveguide into the microring. (e)~FDTD-simulated TE\textsubscript{00} mode of
the ring waveguide at \SI{635}{\nano\meter}, shown for comparison with (d).}
\label{fig:opt_setup}
\end{figure}
Optical access to the fabricated devices is provided by a home-built end-fire
coupling setup, illustrated schematically in Fig.~\ref{fig:opt_setup}(a)--(b).
Figure~\ref{fig:opt_setup}(a) shows the device geometry addressed by the
setup: a bus waveguide with a microring coupled along its length, contacted at
each chip edge by a lensed input and output fiber. Light from a
\SI{635}{\nano\meter} source is passed through an in-line fiber polarization
controller (PC) to set the input polarization state before being launched
into the chip facet via a lensed fiber mounted on a three-axis piezoelectric
micropositioner; the transmitted signal is collected at the output facet by a
second lensed fiber and directed to a calibrated silicon photodetector (PD),
with the electrical output digitized and read out on an oscilloscope
[Fig.~\ref{fig:opt_setup}(b)]. Coupling alignment at both facets, and guiding
along the chip, are monitored throughout via a CMOS camera imaging the chip
from above.

Figure~\ref{fig:opt_setup}(c) documents the three stages of this alignment
process for a representative device: light launched at the input facet
(``Fiber-In''), scattered light visible along the length of the bus waveguide
during propagation (``Coupling: Bus Wg''), confirming single-mode guiding
through the FBMS-written section, and light recollected at the output fiber
(``Fiber-out''). Figure~\ref{fig:opt_setup}(d) shows a top-view scattering
image of the bus waveguide in the vicinity of a microring, in which scattered
light is visible circulating around the ring rim in addition to the bus,
providing direct visual confirmation of evanescent bus-to-ring coupling.
Figure~\ref{fig:opt_setup}(e) shows the corresponding FDTD-simulated
electric field profile of coupled mode into the ring resonator at $\sim\SI{635}{\nano\meter}$,
included for comparison with the coupling geometry observed in
Fig.~\ref{fig:opt_setup}(d).

For the device characterized in Fig.~\ref{fig:opt_setup}, an input power of
$P_{\mathrm{in}} = \SI{1}{\milli\watt}$ launched into the input facet yields a
transmitted output power of $P_{\mathrm{out}} \approx \SI{30}{\nano\watt}$ at
the output facet.
This figure represents the combined contribution of coupling loss at both facets and propagation loss along the 7-mm waveguide length; the present measurement does not decompose these contributions, and doing so - by cutback measurement on matched waveguide arrays or by extraction of the
intrinsic linewidth of the microring resonance - is identified as a direction
for further characterization in Section~\ref{sec:discussion}.

\begin{figure}[htbp]
\centering
\includegraphics[width=0.9\linewidth]{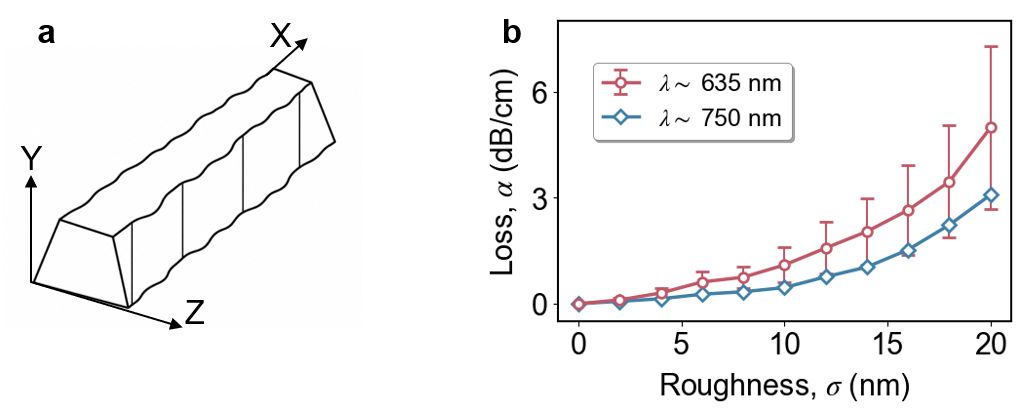}
\caption{Sidewall-roughness-dependent scattering loss model.
(a)~Schematic of a waveguide segment with a synthesized rough sidewall profile
of roughness amplitude $\sigma$, following the exponential autocorrelation
description of Ref.~\cite{roberts2022}. 
(b)~Simulated scattering loss as a function of roughness amplitude $\sigma$
at \SI{635}{\nano\meter} and \SI{750}{\nano\meter}, $\sigma$ swept from 0 to
\SI{20}{\nano\meter}; error bars indicate the spread across independently
generated roughness realizations of the same statistical parameters.}
\label{fig:roughness}
\end{figure}
\subsection{Sidewall-Roughness-Dependent Scattering Loss}
\label{sec:results-roughness}

To relate the observed facet and sidewall morphology
(Section~\ref{sec:results-sem}) to its expected impact on optical loss, a
sidewall roughness model is constructed following the description of waveguide
sidewall statistics ~\cite{roberts2022}. The
sidewall height deviation $h(x)$ along the propagation direction is treated as
a stationary random process characterized by an exponential autocorrelation
function,
\begin{equation}
R(\tau) = \langle h(x)\,h(x+\tau) \rangle = \sigma^{2}\exp\!\left(-\frac{|\tau|}{L_{c}}\right),
\label{eq:acf}
\end{equation}
where $\sigma$ is the RMS roughness amplitude and $L_{c}$ the correlation
length characterizing the lateral extent over which the sidewall profile
remains self-correlated. A rough sidewall profile generated from
Eq.~\eqref{eq:acf} is illustrated schematically in Fig.~\ref{fig:roughness}(a)
and incorporated into a simulated waveguide section for a range of $\sigma$
values. 

Figure~\ref{fig:roughness}(b) shows the resulting scattering loss as a
function of $\sigma$, swept from 0 to \SI{20}{\nano\meter}, at
\SI{635}{\nano\meter} and \SI{750}{\nano\meter}. Loss increases monotonically
with $\sigma$ at both wavelengths, consistent with the $\sigma^{2}$ dependence
of scattering loss on roughness amplitude predicted by the Payne--Lacey
description of sidewall scattering~\cite{payne1994,roberts2022}, and is
systematically lower at \SI{750}{\nano\meter} than at \SI{635}{\nano\meter}
across the sampled range. This trend is consistent with the reduced
sensitivity to sidewall roughness expected at longer wavelength for a fixed
waveguide cross-section, given the $\sigma^{2}/d^{4}$ scaling introduced in
Section~1.

This model does not constitute a direct measurement of the roughness of the
fabricated devices; it provides the expected relationship, via
Eq.~\eqref{eq:acf}, between sidewall roughness statistics and added scattering
loss for the waveguide geometry used in this work. Considered together with
the smooth, low-roughness sidewalls and facets observed by SEM in
Fig.~\ref{fig:sem_facets}, which qualitatively fall toward the low end of the
$\sigma$ range modeled here, this relationship supports the expectation that
the diamond-scribe singulation and FBMS writing processes do not introduce
substantial added scattering loss, without constituting a quantitative
measurement of that loss.

\section{Conclusion and Discussion}
\label{sec:discussion}

We have reported a \ce{Si3N4} photonic platform for the visible band built
around two process elements: fixed-beam moving-stage electron-beam
lithography for stitch-free writing of cm-scale single-mode
waveguides, and pen-type diamond-scribe singulation for chip-facet formation
without a dicing saw or chemical-mechanical polishing. Scribing and cleaving the Si(100) substrate, guided by lithographically
patterned alignment markers, yields end-facets within \SI{2}{\degree} of the
substrate normal at a singulation yield of \SI{80}{\percent}
(16/20). Structural characterization by SEM shows continuous, stitch-free waveguide and taper geometry and flat, undamaged scribed facets, consistent with the intended outcome of both processes. Light was coupled end-fire into fabricated devices and guided along the full chip length, with evanescent coupling from bus waveguide to microring confirmed by top-view scattering imaging; the output power of $P_{\mathrm{out}} \approx \SI{30}{\nano\watt}$ for $P_{\mathrm{in}} = \SI{1}{\milli\watt}$ at $\sim 635$~nm excitation wavelength. A roughness-dependent scattering loss model, following the sidewall statistics description of Roberts \textit{et al.}~\cite{roberts2022}, indicates that scattering loss remains low in the roughness regime consistent with the observed facet and sidewall quality, though this relationship is not a substitute for a direct loss measurement.

The present results point to two directions for further characterization.
First, the insertion budget reported combines facet-coupling and propagation
contributions; cutback measurement on matched waveguide arrays, or
extraction of the intrinsic quality factor from microring resonance
linewidths, would separate these and represents a natural extension of the
present work. Second, the singulation yield of \SI{80}{\percent} indicates
room for further optimization: we attribute the observed failures to
variability in applied scribing force and to residual stress concentration
at the scribe site, and anticipate that a semi-automated scribing jig
enforcing consistent tip trajectory and force would improve yield further,
particularly at substrate thicknesses above \SI{500}{\micro\meter}.

The process elements reported here - FBMS writing, titanium hard-mask pattern
transfer, pulsed-sonication lift-off, and diamond-scribe singulation - rely on equipment available in a broad range of university-scale cleanroom facilities and require no dicing saw, CMP tooling, or DUV stepper. This modest equipment footprint should make the platform adaptable to other
laboratory-scale \ce{Si3N4} photonics efforts operating in the visible band,
particularly those pursuing on-chip integration with intrinsic emitters or
transferred two-dimensional materials. In such applications, the air-clad
waveguide geometry adopted here preserves optical access to the guided mode
along the full device length - the design constraint motivating this work
from the outset.

\section{Acknowledgments}
P.S. acknowledges support from Prime Minister Research Fellowship (PMRF), Govt. of India. A.K.S. acknowledges the IRCC IIT Bombay for funding support. A.K. acknowledges funding support from the Department of Science and Technology via the grants: SB/S2/RJN-110/2017, ECR/2018/001485, and DST/NM/NS-2018/49. We acknowledge funding support from the National Quantum Mission, an initiative of the Department of Science and Technology, Government of India. A.K. acknowledges the support from ANRF via grant number CRG/2022/001170. We also acknowledge the Industrial Research and Consultancy Center (IRCC); the Centre of Excellence in Nanoelectronics (CEN), IIT Bombay; Fundamental Optics, THz and Optical Nanostructures (FOTON) laboratory at TIFR-Colaba, Mumbai; and Sophisticated Analytical Instrument Facility (SAIF); Centre for Research in Nanotechnology and Science (CRNTS) IIT Bombay for providing access in fabricating photonic chip, sample characterization.
\section{Conflicts of Interest}
The authors declare no conflicts of interest.

\bibliography{references}
\end{document}